\documentclass[preprint2,longabstract]{aastex}

\def\cal#1{{\cal #1}}

\def\m@th{\mathsurround=0pt}
\def\n@space{\nulldelimiterspace=0pt \m@th}
\def\biggg#1{{\mbox{$\left#1\vbox to 20.5pt{}\right.\n@space$}}}

\def\beginenum{\begin{enumerate}}
\def\endenum{\end{enumerate}}
\def\bitem{\begin{itemize}}
\def\eitem{\end{itemize}}
\def\bray{\begin{array}}
\def\eray{\end{array}}
\def\begindoc{\begin{document}}
\def\enddoc{\end{document}}
\def\bq{\begin{equation}}
\def\eq{\end{equation}}
\def\bqy{\begin{eqnarray}}
\def\eqy{\end{eqnarray}}
\def\bqyn{\begin{eqnarray*}}
\def\eqyn{\end{eqnarray*}}
\def\bc{\begin{center}}
\def\ec{\end{center}}
\def\bfll{\begin{flushleft}}
\def\efll{\end{flushleft}}
\def\bflr{\begin{flushright}}
\def\eflr{\end{flushright}}
\newcommand{\Avec}{\mbox{\boldmath $A$}}
\newcommand{\Bvec}{\mbox{\boldmath $B$}}
\newcommand{\Evec}{\mbox{\boldmath $E$}}
\newcommand{\Fvec}{\mbox{\boldmath $F$}}
\newcommand{\Gvec}{\mbox{\boldmath $G$}}
\newcommand{\Rvec}{\mbox{\boldmath $R$}}
\newcommand{\Uvec}{\mbox{\boldmath $U$}}
\newcommand{\Vvec}{\mbox{\boldmath $V$}}
\newcommand{\evec}{\mbox{\boldmath $e$}}
\newcommand{\jvec}{\mbox{\boldmath $j$}}
\newcommand{\kvec}{\mbox{\boldmath $k$}}
\newcommand{\nvec}{\mbox{\boldmath $n$}}
\newcommand{\uvec}{\mbox{\boldmath $u$}}
\newcommand{\vvec}{\mbox{\boldmath $v$}}
\newcommand{\wvec}{\mbox{\boldmath $w$}}
\newcommand{\xvec}{\mbox{\boldmath $x$}}
\newcommand{\omegavec}{\mbox{\boldmath $\omega$}}
\newcommand{\Omegavec}{\mbox{\boldmath $\Omega$}}

\usepackage{spr-astr-addons}    

\usepackage{color}

\RequirePackage{color}
\def\imagei{\centerline{\color[gray]{.75}\rule{\hsize}{4pc}}}%
\def\imageii{\centerline{\color[gray]{.75}\rule{4pc}{4pc}}}%

\newcommand{\vdag}{(v)^\dagger}
\newcommand{\emaila}{authors@email.com}

\begin{document}

%

\title{On the Relaxed States in the Mixture of Degenerate and Non-Degenerate Hot
Plasmas of Astrophysical Objects}

\shorttitle{<On the Relaxed States in a Mixture of Relativistic Plasmas>}
\shortauthors{<Shatashvili et al>}

\author{ N.L. Shatashvili\altaffilmark{1,2}}
\altaffiltext{1}{Andronikashvili Institute of Physics, TSU,
Tbilisi 0177, Georgia} \
\altaffiltext{2}{Department of Physics,
Faculty of Exact and Natural Sciences, Ivane Javakhishvili Tbilisi
State University (TSU), Tbilisi 0179, Georgia}
\author{ S.M. Mahajan\altaffilmark{3}}
\altaffiltext{3}{Institute for Fusion Studies, The University of
Texas at Austin, Austin,Tx 78712} \and
\author{V.I. Berezhiani\altaffilmark{1,4}}
\altaffiltext{1}{Andronikashvili Institute of Physics, TSU,
Tbilisi 0177, Georgia}
\altaffiltext{4}{School of Physics, Free
University of Tbilisi, Georgia} \

\begin{abstract}
It is shown that a small contamination of a relativistically
hot electron component can induce a new scale (for structure formation)
to a system consisting of an ion-degenerate electron plasma.
Mathematically expression of this additional scale length is
the increase in the index of quasi-equilibrium Beltrami-Bernoulli states
that have been invoked to model several astrophysical systems of interest.
The two species of electrons, due to different origin of their relativistic
effective masses, behave as two distinct components (each with its own
conserved helicity) and add to the richness of the accessible
quasi equilibrium states. Determined by the concrete parameters
of the system, the new macro-scale lengths (much larger than the
short intrinsic scale lengths (skin depths) and generally much
shorter than the system size) open new pathways for energy transformations.
\end{abstract}


\maketitle


\section{Introduction}

In many compact astrophysical objects, the plasma density
is so high that the mean inter-particle distance becomes
smaller than the De Broglie thermal wavelength. The Fermi
Energy for such a highly degenerate gas (obeying Fermi-Dirac
statistics) can become relativistic, especially for the
lighter electrons. Consequently, the degeneracy pressure
may easily dominate the thermal one
[see e.g. \citep{Compact-1}, \\
\citep{Compact-2,BSM_deg}, \\
\citep{degenerate,SMB_multi} and references therein].
Relativistic outflows/jets, often, come out of these
compact objects. The interaction of the Relativistic ejecta
with their degenerate shells could create turbulence and
shock waves at short scales. On the large scale, however,
one could still expect the interacting system to be describable
in terms of quasi equilibrium states. An interesting
representative of such systems is a White Dwarf (WD) --
the most common endpoint of stellar evolution
\citep{White,hmfwd},\\
\citep{schmidt,Kawka},\\
\citep{kulebi,kepler},\\
\citep{DAZ,hollands}.

Since many stars are born in binary systems with
sufficiently small initial separations, they go through one
or more phases of mass-exchange (see e.g. \\
\citep{winget,tremblay}, \\
\citep{mukai} and references therein). The solar neighborhood
is populated by numerous accreting white dwarfs (AWDs) that are surrounded
by an accretion gas of companion star or disk \citep{Begelman,mukai}.
The composite system is a highly interesting and unusual state of
matter; a highly degenerate WD plasma co-existing
with a classical hot accreting astrophysical flow.

Recently it was shown that accreting WD atmospheres may support
degenerate plasma relaxed states \citep{BSM_deg} that have
associated fast super-Alfv\'enic up-flows with dramatically
reduced densities \citep{BS-flow}. It is expected that this
combination -- a bulk degenerate plasma contaminated by small fraction of
a non-degenerate highly relativistic plasma -- will also pertain during the
relativistic jet formation from accretion-induced collapsing White Dwarfs
to Black Holes \citep{Begelman,Kryvdyk,JetsWD}.

\bigskip

Following the conceptual framework developed in the study of
multi-component Relaxed States (including relativistic ones)
[see e.g. \citep{MY,iqbal-1,mmns-1}, \\
\citep{relaxed,DB},\\
\citep{morrison,ymois}, \citep{iqbal-3,pino,Multi-B}, \\
\citep{dewar2,SMB_multi} and references therein],
we will study, in this paper, the quasi equilibrium states
that are accessible to the system composed of two electron
species (a highly degenerate main component (to be called $d$)
mixed with a smaller classical relativistic flow (to be called $h$))
immersed in a neutralizing ion background; the latter could
be either static or dynamic. One of the principal aims of such
a search is to extract any new scales of structure formation
induced by the addition of new physics to the system. One will,
thus, look for the $h$ component induced, intermediate macroscopic
length scale $L_{macro}$ [$L_{meso}$] lying between the system
size and the relatively small intrinsic scales (measured by the
skin depths); a knowledge of these scales may help us
better understand the evolution of accreting stars.

\section{Model Equations}

We will be studying a quasi neutral plasma of a mobile classical
ion component ($i$), and two relativistic electron species --
the bulk $d$ electron gas with a density $N_{0d}$ and a small
contamination of $h$ electrons with density $N_{0h}$.
The quasi neutrality demands
\begin{equation}
N_{0d} + N_{0h} = N_{0i} \ \ \Longrightarrow \ \
\frac{N_{0i}}{N_{0d}} = 1+\alpha, \quad   \alpha \equiv \frac{N_{0h}}{N_{0d}} \  ,
\label{B-eq}
\end{equation}
where \ $\alpha \ll 1$ \ labels the ratio of hot electron fraction.
to the degenerate electrons.

The electron dynamics for both components will be described by the
appropriate relativistic fluid equations [\citep{pino}, \citep{BSM_deg}, \\
\citep{degenerate}]:
the  continuity
\begin{equation}
\frac{\partial N_{d(h)}}{\partial t} \ + \ \nabla \cdot ({N_{d(h)}{\bf
V}_{d(h)}})=0 \ , \label{Cont}
\end{equation}
and the equations of motion:
\[
\frac{\partial}{\partial t}\left(G_{d(h)}\,{\bf p}_{d(h)}\right) \ + \
m_ec^{2}\, {\bf \nabla}\left(G_{d(h)}\ \gamma_{d(h)}\right)
\]
\begin{equation}
= - \ e\,{\bf E} \ + \ {\bf V_{d(h)}}\times \Omega_{d(h)} \
\label{B6}
\end{equation}
\noindent where  \ ${\bf p}_{d(h)}= \gamma_{d(h)} m_e{\bf
V}_{d(h)}$ \ is the hydrodynamic momentum, \ $n_{d(h)} =
N_{d(h)}/\gamma_{d(h)}$ \ is the rest-frame particle density \
($N_{d(h)}$ \  denotes the laboratory frame density) \
of the degenerate (hot) electron fluid element, \
${\bf V}_{d(h)}$ \ is the fluid velocity, and
$\gamma_{d(h)} =\left( 1-V_{d(h)}^{2}/c^{2}\right)^{-1/2}$ \ .

Notice that the factors $G_{d}$ and $G_{h }$, constituting
what could be seen as the effective mass, are quite
different for the two electron species. In particular,
\ $G_{d} = w_{d}/n_{d}m_{e}c^{2}$ , \ where \ $w_{d}$ \ is
an enthalpy per unit volume, originates from degeneracy
rather than relativistic kinematics. The general expression for enthalpy
\ $w_{d}$ \ for arbitrary density and temperature (for a plasma
described by local Dirac-Juttner equilibrium distribution
function) can be found in \citep{Russo-2}. For a fully (strongly)
degenerate electron plasma, however, this very
tedious expression smoothly transfers to the one with just density
dependence: $w_{d}\equiv w_{d}(n)$ \citep{BSM_deg}. In
fact \ $w_{d}/n_{d}m_{e}c^{2}=\left(
1+(R_{d})^{2}\right)^{1/2}$, where \ $R_d$ [$=
(n_d/n_c)^{1/3}$ \ with \ $n_c = 5.9\times 10^{29}cm^{-3}$
being the critical number-density]. The effective mass factor,
then, is simply determined by the plasma rest frame density,
$G_{d}=[ 1+(n_{d}/n_{c})^{2/3}]^{1/2}$ \ for arbitrary \ $n_{d}/n_c$ .
For relativistically hot plasma an expression for effective mass
factor $G_h$ can be found in \citep{BM-94,Ryu}.

On taking the {\it curl} of these equations, one can cast them
into an ideal vortex dynamics ((Mahajan 2003, 2016) and references
therein)
\begin{equation}
\frac{\partial}{\partial t}\,{\bf \Omega}_{d(h)}={\bf
\nabla\times}\left( {\bf V_{d(h)}\times\Omega_{d(h)}}\right) \ ,
\label{B7}
\end{equation}
in terms of the generalized (canonical) vorticities \  ${\bf
\Omega_{d(h)}=} - (e/c ) {\bf B+\nabla\times}\left( G_{d(h)}{\bf
p}_{d(h)}\right)$ . Note that  generalized vorticity
would acquire an additional term for isentropic systems [see \citep{M-EV}];
the present derivation pertains only to the homentropic plasmas.
For special astrophysical conditions, canonical vorticity
would also have a quantum-mechanical part [see \citep{SpinVort,Felipe}
for spinning plasmas], even a general relativistic component
[see e.g. \citep{AccrVort} for Black Hole accretion disks]
in addition to the electromagnetic, kinetic and thermal contributions.
For most applications these corrections are negligibly small.
Numerical estimates of spin-magnetic interactions of two-fluid
plasmas of white dwarfs and neutron stars was presented in
\citep{Gomez}, and was shown to be not so consequential.

\bigskip

Ion fluid dynamics is described by the corresponding Continuity
Equation and the following Equation of Motion:
\[
m_i\left[\frac{\partial {\bf V}_i}{\partial t} + ({\bf V}_i\cdot
\nabla){\bf V}_i\right] = -\frac{1}{N_i}\nabla p_i \ +
\]
\begin{equation}
\hspace{3.8cm} + \ e{\bf E} \ + \frac{e}{c}\,{\bf V}_i\times {\bf B} \ .
\label{B6i1}
\end{equation}

\bigskip

The low frequency dynamics is, now, closed with Ampere's law
\begin{equation}
\nabla \times {\bf B} =\frac{4\pi e }{c}\, \left[ (1 + \alpha)\,N_i
\,{\bf V}_i - N_d\,{\bf V}_d - \alpha \,N_h\,{\bf V}_h\right] ,
\label{B-Amp1}
\end{equation}
another relation between \ ${\bf V}_i \ , {\bf V}_{d(h)}$ \ and \
${\bf B}$. Notice that the small hot electron population,
represented by \ $\alpha$ \ and \ ${\bf V}_h$,
will become the source of a new scale-length; Finding and exploring
this scale length (which adds the diversity to the scale-hierarchy
of multi-component plasmas met in astrophysical conditions)
is the principal objective of this paper.

\bigskip

We will concentrate  on a special class of equilibria known as the
Beltrami-Bernoulli (BB) states \citep{BSM_deg}. We expect to find
the new channel for energy transformations in such a mixture
of relativistic plasmas often emerged while the evolution of
certain astrophysical objects, specifically while the evolution of
accreting stars; star collapsing and etc..

\section{Equilibrium States in 2-temperature relativistic degenerate electron-ion Plasma}

In this paper, the density is normalized to \ $N_{0d}$ (the
corresponding rest-frame density is \ ${n_{0d}}$); the magnetic
field is normalized to some ambient measure $|{\bf B}_0|$; hot electron
gas temperature is normalized to $m_ec^2$; all
velocities are measured in terms of the corresponding Alfv\'en
speed \  $V_A = V_{Ad} = B_0/\sqrt{4\pi n_{0d}m_eG_{0d}}$ ; all
lengths [times] are normalized to the "effective" degenerate electron skin skin
depth \ $\lambda_{\rm{eff}} \, [\lambda_{\rm{eff}}/V_A]$ , where
\begin{equation}
\lambda_{\rm{eff}} \equiv \lambda_{\rm{eff}}^{d}
=
\frac{c}{\omega_{pe}^d}=c\,\sqrt{\frac{m_eG_{0d}}{4\pi n_{0d}e^2}} \
. \label{lamb}
\end{equation}
Notice, that "effective" degenerate electron skin depth is
related to the "effective" hot electron skin depth $\lambda_{\rm{eff}}^h$ as
\begin{equation}
\lambda_{\rm{eff}}^d =
c\,\sqrt{\frac{\alpha \,G_{0d}}{G_{0h}}}\ \sqrt{\frac{m_eG_{0h}}{4\pi n_{0h}e^2}}
=\sqrt{\frac{\alpha \,G_{0d}}{G_{0h}}} \ \lambda_{\rm{eff}}^h \
, \label{lambd}
\end{equation}
where
\[
\lambda_{\rm{eff}}^h = c\,\sqrt{\frac{m_eG_{0h}}{4\pi n_{0h}e^2}} \ .
\]
Depending on $\alpha $, and the degeneracy level as well as the
relativistic temperature of the fraction (outflow/jet), there are
2 drastically different length scales in addition to the conventional
ion-skin depth,
\begin{equation}
\lambda_{i} = \frac{c}{\omega_{pi}}=c\,\sqrt{\frac{m_i}{4\pi n_{0d}e^2}} \
. \label{lambI}
\end{equation}
Here
\begin{equation}
G_{0d}(n_{0d}) = [1 + (R_{0d})^2]^{1/2}
\label{G0}
\end{equation}
with
\begin{equation}
\qquad \qquad R_{0d}=\left( \frac{n_{0d}}{n_c}\right)^{1/3};
\label{R0}
\end{equation}
while \citep{mignone,Ryu}
\begin{equation}
G_{0h}= \frac{5}{2}\frac{T_{e0}}{m_ec^2}+\frac{3}{2}\sqrt{\left(\frac{T_{e0}}
{m_ec^2}\right)^2+\frac{4}{9}} \ .
\label{H0}
\end{equation}
One must emphasize  that the intrinsic skin depths (the natural
length scales of the dynamics), though over a large length span,
are much shorter compared to the system size. For $d$ electrons,
the effective mass goes from \ $G_{0d}(n_{0d}) = 1 +
\frac{1}{2}(\frac{n_{0d}}{n_c})^{2/3}$ \ in the
non-relativistic limit ($R_{0d}\ll 1$ ) to  \ $G_{0d}(n_{0d}) =
\left(\frac{n_{0d}}{n_c}\right)^{1/3}$ in the ultra-relativistic
regime ($ R_{0d}\gg 1$ ), and for the $h$ component,
the effective mass goes from $G_{0h}(T_{e0})=
1+\frac{5}{2}\frac{T_{e0}}{m_ec^2}$ \ in the
non-relativistic limit ($T_{e0}\ll m_ec^2$ ) to
\ $G_{0h}(T_{e0})=4\,\frac{T_{e0}}{m_ec^2}$
in the ultra-relativistic regime ($ T_{e0}\gg m_ec^2$ ).

\bigskip

By following the methodology of \citep{pino} and
\citep{BSM_deg,SMB_multi}, we obtain the BB equilibrium conditions
for both the $d$ and $h$ electrons (the primary difference is in
the physics of $G_{d}$ and $G_h$). The Beltrami conditions
\begin{equation}
{\bf B} - \nabla \times (G_{d}\,\gamma_{d}{\bf V}_{d}) =
\ a_{d}\,\frac{n_{d}}{G_{d}}\,(G_{d}\,\gamma_{d}{\bf
V}_{d}) \ , \label{B9}
\end{equation}
\begin{equation}
{\bf B} - \nabla \times (G_{h}\,\gamma_{h}{\bf V}_{h}) =
\alpha \ a_{h}\,\frac{n_{h}}{G_{h}}\,(G_{h}\,\gamma_{h}{\bf
V}_{h}) \ , \label{B9h}
\end{equation}
align the Generalized vorticities along their velocity fields.
This Beltrami alignment imposes (on the electron fluids)
the following generalized  {\it Bernoulli Conditions},
expressing the balance of all remaining potential forces,
 \begin{equation}
\qquad \qquad \nabla (G_{d}\,\gamma_{d}\ - \ \varphi )=0 \quad
\label{B8}
\end{equation}
and
\begin{equation}
\qquad \qquad \nabla (G_{h}\,\gamma_{h}\ - 
\varphi )=0 \quad
\label{B8h}
\end{equation}
may be combined to form
\begin{equation}
\qquad \qquad G_{d} \,\gamma_{d} + G_{h}\,\gamma_{h}  -
2 \,\varphi = const \ ,
\label{Bernoulli}
\end{equation}
where $\varphi $ is the electrostatic potential
(of purely electromagnetic nature).
This set, coupled with Ion fluid Beltrami Condition:
\begin{equation}
{\bf B} + \zeta \nabla \times {\bf V}_i = (1+\alpha) \,a_i n_i{\bf
V}_i , \quad  \zeta = \left[G_0^d\,
\frac{m_e^d}{m_i}\right]^{-1} \ , \label{IBC}
\end{equation}
together with Ampere's law  Eq.(\ref{B-Amp1}), defines the BB
equilibrium states pertinent to the system of a two electron
component ($d$ and $h$) fluid immersed in a neutralizing ion fluid.
The separation (proportionality) constants \ $a_{d(h),i}$ \
are related to the system invariants, the total energy, and the
{\it generalized helicities} for each component,
\begin{equation}
h_{d(h),i}=\int ( {curl^{-1}{\bf \Omega}_{d(h),i}) \cdot {\bf
\Omega}_{d(h),i}\,d {\bf r} } \ . \label{B10}
\end{equation}
Below we put $\varphi \equiv 0$ \ due to the quasi-neutrality hold
throughout the overall dynamics assuming the incompressibility;
gravity will be ignored for the time being.

The asymmetry between the bulk electron ($d$) and the ion fluid is
due to a small fraction of hot electrons ($\alpha\ll
1 , \ |{\bf V}_i|\ll|{\bf V}_{d(h)}| $). Notice that there are,
in fact, two symmetry-breaking mechanisms in this model:
1) the first is due to different effective inertias for the $d $
and $h$ electrons, and 2) the second is from the  small h contamination
added to the bulk $d$ electrons ($\alpha \neq 0 , {\bf V}_h \neq 0$).
Each one of these is responsible for creating a net ``current''. The
structure formation mechanism explored in \citep{structuresPI-1,structuresPI-2},
\citep{DB}, \citep{mmns-1}, \\
\citep{mnsy}, \citep{osym}), originates, for instance, in the effective inertia
difference. Asymmetry between the plasma constituents increases
the number of conserved helicities, and eventually translates into
a higher index Beltrami state \citep{Multi-B,SMB_multi}. It should also
be mentioned that due to the different origin of relativistic
effective masses \ $G_d \neq G_h$, the index of the Beltrami system
is determined by the simultaneous action of both asymmetries.

\section{Quadrupole Beltrami Equations}

In this section we show that an appropriate but tedious manipulation
of the set of the Eq.-s (\ref{B-Amp1})-(\ref{IBC}), leads us to
an explicit quadruple Beltrami equation. The variable of choice
turns out to be the Hot electron Fluid Velocity ${\bf V}_h$
(the Beltrami index is measured by the highest number of {\it curl}
operators \citep{Multi-B}).

\bigskip

From Ampere's law in dimensionless form variables as:
\begin{equation}
\nabla \times {\bf B} = 
\left[ (1+\alpha )\,
\frac{N_i}{N_{0i}}{\bf V}_i -\frac{N_d^e}{N_{0d}^e}{\bf
V}_d - \alpha \,\frac{N_h^e}{N_{0h}^e}{\bf V}_h) \right] \label{B-Amp},
\end{equation}
we find that if \ $\alpha=0$ \ (no $h$ contamination, then
quasineutrality reads $N_i=N^d=N$), we have:
\[
\qquad \qquad \qquad {\bf V}_d \equiv {\bf V}_e={\bf V}_i -
\frac{1}{N}\ \nabla \times {\bf B}
\]
that will  reproduce the Double Beltrami (DB) state, relevant to an
ion-degenerate electrons plasma \\
\citep{BSM_deg}. For $\alpha \neq 0$ \ and \ ${\bf V}_i=0 $
(immobile ions), the relation
\[
\qquad \qquad \qquad {\bf V}_d = - \frac{N_h}{N_d}\,{\bf V}_h
- \frac{1}{N_d^e}\ \nabla
\times {\bf B}
\]
will lead to higher (Triple) Beltrami states when inertia effects in the
degenerate and hot electron fluids are taken into account.

Observations show that hot electron fluid fraction can be small ($\alpha
\ll 1$); ion fluid velocity are also much smaller than those for
lighter electron ($d$, $h$) fluids \ [${\bf V}_i \ll
{\bf V}_d , {\bf V}_h $].  Thus, ion dynamics could be
neglected in most of cases \citep{relaxed} except for $\alpha = 0$ when
flow effects can be crucial in creating the structural richness
in astrophysical environments, in the heating/cooling processes,
and in Generalized Dynamo theory and flow acceleration phenomena
\citep{mmns-1,mnsy,msms}, \citep{lmRD}.

For $\alpha = 0$ (pure e-i plasma with degenerate electrons),
it was shown that when electron inertia is neglected, the system
reduces to a Double Beltrami state \citep{BSM_deg}. One could expect,
then, that for the full model described in this paper, the composite
Beltrami condition of index 4 will arise as two distinct "components"
are  being added. The index would fall to three when the
ion flow effects are neglected (${\bf V}_i\to 0$).

\bigskip

Let us now explore the new structures accessible to the Beltrami
states for the full model -- the $d$ and $h$ electrons and mobile ions.
We will assume $\varphi \equiv 0$, make the simplifying assumption
\ $\gamma_d \equiv 1 , \ \ \gamma_h\equiv 1$ \ that reduces the
Bernoulli Conditions (\ref{B8},\ref{B8h}) to \ $G_d = const ; \ \ G_h = const$ .
The resulting Ampere's law (\ref{B-Amp}), in dimensionless variables,  becomes
\begin{equation}
\qquad \qquad \nabla \times {\bf B} = [(1+\alpha) \,{\bf
V}_i - {\bf V}_d \ - \alpha {\bf V}_h] \ .
\label{B11}
\end{equation}
In terms of the bulk "Flow velocity" (combining $d$ electrons and ions),
\begin{equation}
\qquad \qquad {\bf V} = \frac{1}{2}\,[(1+\alpha)\,{\bf V}_i + {\bf V}_d]
\label{B12'}
\end{equation}
one can express {\it the Generalized ion Velocity and Momentum for
$d$ electrons} as [$ {\bf P}_{d}^e=G_{0d}^e(n_{0d}^{e})\,{\bf V}_{d}^e \ $] :
\begin{equation}
{\bf V}_{i} = \frac{1}{1+\alpha}\,\left({\bf V} \ + \ \frac{1}{2}\,\nabla \times
{\bf B}+\frac{\alpha}{2}\,{\bf V}_h \right) \ ,
\label{B13'}
\end{equation}
\begin{equation}
{\bf P}_{d} = G_{0d}^e\,\left({\bf V} \ - \ \frac{1}{2}\, \nabla \times {\bf
B}-\frac{\alpha}{2}\,{\bf V}_h \right) \ ,
\label{B13}
\end{equation}
and  the Ion flow Beltrami condition ({\ref{IBC}) as
 \begin{equation}
\qquad \qquad \qquad {\bf B} + \zeta \,\nabla \times {\bf V}_i =
(1+\alpha ) \,a_i \,{\bf V}_i \ .
\label{IBC-2}
\end{equation}
Straightforward algebra, using Eqs. (\ref{B13'}) and (\ref{B13}) in
Eqs. (\ref{B9}) and (\ref{IBC-2}), leads to ($G_{0d}(n_{0d})\equiv G_0$):
\[
{\bf V} = \eta \left( \beta \,\nabla
\times \nabla \times {\bf B} - \frac{1}{2}\,[a_i(1+\alpha)^2\,\beta - a_d]\,
\nabla \times {\bf B}\right) \
\]
\[
\qquad \ + \ \eta \,[1 + \beta \,(1+\alpha) ]\,{\bf B}  \ + \ \alpha \, \beta \, \,
\nabla \times {\bf V}_h \ - \
\]
\begin{equation}
\qquad \ - \ \frac{\alpha}{2}\,[a_i(1+\alpha)^2\,\beta - a_d]\,{\bf V}_h
\label{B14}
\end{equation}
\[
{\rm with} \quad \eta \,\equiv [a_i(1+\alpha )^2\,\beta + a_d]^{-1}
\quad , \quad \beta \equiv G_0\,\zeta^{-1} \ .
\]
The parameter $\beta$ is a measure of degeneracy as well as the mobility
of ions: $\beta \to 0$ for immobile ion fluid \ ($m_i \to \infty $) \ \
and \ $\beta \ll 1, \ R_{0d} \ll1$ \ [$\beta > 1, \ R_{0d} \gg 1$] \ \ for
the weakly [strongly] degenerate electrons; in the latter case,
the degenerate electron fluid inertia can not be ignored. In the limit
of a pure e-i plasma \ ($\alpha \equiv 0; \ \ \beta \ll 1; \ \ \eta
\simeq a_d^{-1}$) \ , the pertinent simple relation \ ${\bf V} - \frac{1}{2}
\ \nabla \times {\bf B} = a_d^{-1}{\bf B}$ \ reveals that the inertialess
electrons move parallel to magnetic field.

\bigskip

Further manipulation of the system is displayed in Appendix A, the
end result is the emergence of the quadruple Beltrami equation (QB)
for ${\bf V}_h$ for arbitrary $\alpha $ [$G_h \equiv G_{0h} = H_0]$:
\[
\eta \,G_0\,\nabla \times \nabla \times \nabla \times \nabla
\times {\bf V}_h + \eta \,G_0 b_1\,\nabla \times \nabla \times
\nabla \times {\bf V}_h
\]
\begin{equation}
+ \eta \,b_2\,\nabla \times \nabla
\times {\bf V}_h - b_3\,\nabla \times {\bf V}_h -2\alpha \,b_4\,{\bf V}_h =
0 \ . \label{QB-1}
\end{equation}
Naturally such a system will be endowed with four distinct length
(constructed from the defining parameters). Different effective masses
of the degenerate bulk population and of a hot electron
contamination and their ratio are the new elements of physics
introduced in this paper. Notice that if either $\alpha $ ($h$ fraction)
or the $b_4$ (ion mobility factor) were zero, the Beltrami index
of the system goes down implying the disappearance of a scale length.

Solving the Eq.(\ref{QB-1}) for ${\bf V}_h$ and plugging it
into (\ref{B9h}) we will get the equation for ${\bf B}$; for
the pure incompressible degenerate e-i plasma it is better to use
Eq.(\ref{B-4}) (with $\alpha \equiv 0$) directly to find the
magnetic field \ ${\bf B}$.

\subsection{Assymetry Induced Macroscopic Structure Formation}

Justified by observational evidence (see introduction), we will
assume $\alpha \ll 1$ that will simplify the coefficients in Eq.(\ref{QB-1}).
A formal factorization of ({\ref{QB-1}}}) leads to
\begin{equation}
(curl - \mu_1)(curl - \mu_2)(curl - \mu_3)(curl - \mu_4)\ {\bf
V}_i = 0 \ , \label{Q-curl}
\end{equation}
where the inverse length scales $\mu_i$ are functions of \ $\alpha , \
\beta , \ n_{0d} , \ H_0$ \ and the degeneracy-determined mass factor \
$G_0$ . The general solution of Eq.(\ref{Q-curl}) is a sum
of four Beltrami fields ${\bf F_k}$ (solutions of Beltrami
Equations $\nabla \times {\bf F}_k = \mu_k {\bf F}_k$) \ while
eigenvalues ($\mu_k $) of the {\it curl} operator are the
solutions of the fourth order equation
\begin{equation}
\qquad \qquad \mu^4-b_1^{*}\,\mu^3 + b_2^{*}\,\mu^2 - b_3^{*}\,\mu
+ b_4^{*} = 0 \ .
\label{mu-eq}
\end{equation}
Details of a similar analysis can be found in \\ \citep{SMB_multi}.
The interesting and important result of this enquiry follow after an
examination of the various $b^{*}$ coefficients of (\ref{mu-eq}).

Though the inverse scales, determined by \ $b_1^{*}$, $b_2^{*}$, and
$b_3^{*}$ , do get slightly modified by $\alpha \ll 1$
corrections, it is the inverse scale associated with \ $b_4^{*}$ \
that is most profoundly affected; being proportional to $\alpha$,
it tends to become small, i.e, {\it the corresponding scale length
becomes large} as $\alpha$ approaches zero; the corresponding scale length
becomes strictly infinite for \ $\alpha = 0$, and disappears
reducing (\ref{mu-eq}) to a triple Beltrami system.

Thus the asymmetry induced due to the small fraction of
relativistically hot electrons may lead to the formation of
macroscopic structures through creating an intermediate/large
length scale, much larger than the intrinsic scale skin depths
(but less than the system size). It is important to note that
this mechanism operates for all levels of bulk electron as long as
degeneracy (the range of \ $G_0$ \ was irrelevant) and the hot
electron fraction is nonzero. The possible significance and importance of
natural mechanisms of this sort (such cases are natural in astrophysical
conditions as discussed in the introduction) for creating
Macro-structures in astrophysical objects was already
discussed in \citep{SMB_multi} for different type asymmetric multi-fluid
systems.

\section{Scale Hierarchy in 2-temperature relativistic e-i plasmas}

The new macroscopic scale discussed in previous section
can be ``determined" by dominant balance arguments: As the
scale gets larger, \ $|\nabla|$ \ gets
smaller, and the dominant balance will be between the last terms
of (\ref{mu-eq}), yielding [we remind the reader, that all lenghts
are normalized to the \ $\lambda_{\rm{eff}}$ , and \ $\zeta \gg 1
$ \ even for ultra-relativistic case]:
\begin{equation}
\qquad \qquad L_{\rm{macro}}= {\frac{1}{2\alpha}}\,{\frac{|b_3|}{|b_4|}} =
\frac{C}{2\alpha} \label{L_macro}
\end{equation}
where $C(a_i , a_d , a_h , G_0 , H_0 , \beta )$
is a rather complicated function of the plasma-system parameters
(see Appendix A for $b$-coefficients).

Let us assume that the densities of the 2-temperature relativistic
electron-ion plasmas of interest are such that $(1-\beta) \gg \alpha$
( $\beta \leq 1$ even for the ultra-relativistic case when the
degenerate electron component density range is within \ $(10^{25}
- 10^{34})\,cm^{-3}$). We can, then simplify $C$ for
$(a_d/a_i) \leq \beta \ , (a_h/a_i) \leq \beta $ and write
\begin{equation}
\qquad L_{\rm{macro}} \sim \frac{H_0}{\alpha \,[\beta +a_h(1-\beta)]} \ |(1-\beta)| \ .
\label{Lmacro}
\end{equation}

There are 2 possible, observationally relevant, limiting cases:

(i) The degenerate electron fluid density is so high ( $10^{33}
- 10^{34}\,cm^{-3}$, strongly relativistic Fermi energy)
that $\beta$ approaches unity [typically $ 0.1 - 0.5$ ].
The macroscopic length (\ref{Lmacro}), then, yields,
\begin{equation}
\qquad \qquad L_{macro} \sim \frac{H_0}{\alpha } \
\frac{1-\beta}{\beta} \gg 1 \ .
\end{equation}

(ii) The degenerate electron fluid density is in the lower range
($\sim 10^{25} - 10^{32}\,cm^{-3}$) leading to a $\beta \ll 1$.
The  expression for $L_{\rm{macro}}$ simplifies to
\begin{equation}
\qquad \qquad L_{\rm{macro}} \geq \frac{H_0}{\alpha} \
\frac{1}{a_h} \gg 1 \
\end{equation}
for any $H_0 > 1$ and $G_0> 1$.

Notice that the hot electron induced $L_{\rm{macro}}$ for
the strongly degenerate bulk electrons tends to be smaller
than the corresponding length for low bulk degeneracy.

\bigskip

It is important to consider another obviously interesting
 \ $\beta \ll 1$ \ (weakly degenerate) case when
$a_d \sim a_h \equiv a$ and they are both $\sim a_i$
(note that under the same assumptions \ $\eta \sim a^{-1}$).
The  $b^{*}$-coefficients take the form (reminding that
$\beta^{-1}\,G_0 = \zeta \gg 1$):
\begin{equation}
b_1^{*} = \zeta \,a \ , \quad
b_2^{*} = 2\,H_0\,G_0^{-1}(1 +  0.5\,a^2) +\alpha \,\zeta \,a^2 \ ;
\label{b-s}
\end{equation}
\[
b_3^{*} = 
- \alpha \,a\,[\zeta + G_0^{-1}\,(1 + 0.5\,a^2)] \ ;
\
b_4^{*} = \frac{\alpha}{2}\,G_0\,a^2\,(1-a)  .
\]
We see that depending on the physical parameters: \ $H_0, \ G_0 ,
\ \alpha ,\ \zeta$ , different scale hierarchies will emerge.

Even at a very small fraction of the hot electron fluid
($\alpha \ll 1$), none of $b^{*}$-coefficients
vanish -- they remain finite and the macro/meso scale
is always present in such a system. Then, for $a \gg 1$ ($a\ll 1$),
we will have quadruple (triple) Beltrami states
in our complex relativistic system. When $a \equiv 1$
we have a Triple-Beltrami equilibrium.  If in addition,
the $h$ electron fraction vanishes ($\alpha \equiv 0$), the equilibrium
reduces to a Double-Beltrami state consistent to previous results \citep{BSM_deg}.

The three component plasma (ions, and two species of  electrons),
studied in this paper, is another example of the rule that the
associated BB equilibria follow, i.e, the Beltrami index is \ $I = M+1$ ,
where $M$ is the number of ``independent" components. Naturally the
index is a measure of the independent characteristic scale
lengths [see Mahajan \& Lingam 2015].

\bigskip

The scale hierarchy, pertinent to our model of astrophysical
significance, may be summarized as follows:

(1) For an ion-degenerate electron plasma, the equilibrium
is triple Beltrami with the following fundamental three
scales;  the system size $L$, and the two intrinsic scales,
the $d$-electron and ion skin depths. If $d$-electron inertia
is negligible (relatively lower densities), then equilibrium
collapses to a double Beltrami.

(2) Both the skin depths associated with $d$ and $h$, that
are microscopic in a non degenerate/non-relativistic
plasma, can become larger due to relativistic effects
and could be classified as meso-scales [$l_{\rm{meso}}$].
Under some special constraints on the Bertrami parameters,
the meso-scale $l_{\rm{meso}}$ can become very large!

(3) With the relativistically hot electron ($h$) species
acting as an independent component, the equilibrium becomes
quadruple Beltrami with a new additional scale, $L_{\rm{macro}}$.
Originating entirely in the $h$ fraction ($\alpha \neq 0$),
this scale disappears as this fraction goes to zero.
Both the larger ion mass and lower density hot electron
fraction contribute towards boosting $L_{\rm{macro}}$.

(4) In the limit of immobile ions (only $d$ and $h$ electrons
are the dynamical components), the equilibrium is triple
Beltrami (as expected), and the largest intrinsic scales are
the relativistically enhanced skin-depths of the two species,
the relative scale size will be determined by the ratio $G_0/H_0$.

\section{Summary}

We studied the quasi equilibrium Beltrami-Bernoulli states
that are accessible to a three component plasma composed of two electron
species (a highly degenerate main component mixed with
a smaller classical relativistic hot flow) immersed in a neutralizing
ion background; the latter could be either static or dynamic.
Study of such a plasma could be generally relevant to the
evolution of certain astrophysical objects, specifically
during the accretion stage of stars; star collapsing etc.
The two electron species -- the bulk degenerate electron fluid
($d$) and a small contamination of relativistically hot electrons ($h$)
-- contribute two components to a three component plasma
that in addition, has a neutralizing ion background.

The $h$ contamination has the expected but striking effect of
providing an added  macroscopic scale lying between the system
size and the relatively small intrinsic scales (measured by the
skin depths). The existence of a new scale for structure formation
may provide crucial insights into the evolution of accreting
stars; in particular, new channels for energy
transformations may become available [\citep{mnsy,osym,msms}, \\
\citep{Shiraishi}, \citep{SY-DJ}].

Although not the direct subject matter of this paper,
the existence of a small relativistic electron component
is reminiscent of the runaway electron population in a Tokamak.
It is conceivable that the energy exchange processes studied
in this paper, could have some relevance to runaway induced phenomena.

The new element of physics introduced in this paper arises due
to the different effective masses of the $d$ and $h$ electrons.
In fact, that is what forces us to treat them as independent components.
It is this piece of physics that leads to the creation of the
$L_{\rm{macro}} [l_{\rm{meso}}]$ alluded to on the preceding paragraph.

These macro/intermediate scales, opening new pathways for
energy transformations, can advance our understanding
of a host of quiescent as well as explosive astrophysical phenomena --
magnetic field generation, structure formation, fast/transient
outflow and jet formation, heating/cooling etc. We plan to explore
the consequences of the particular findings of this paper in the
context of the accreting White Dwarf evolution problem, and the
events accompanying the phenomenon of star-collapse.

Finally, when a complex system moves from Double- to Triple-
and Quadruple-Beltrami states, the roots (inverse length-scales)
will exhibit a wide range of behavior. In the quadruple case,
there are possible transitions from 2 complex-conjugate pairs to: 1)
one complex-conjugate pair and two real roots, and 2) to 4 real roots.
In each of these cases, the conversion of magnetic energy into flow energy
can occur; the process will result in energy transfer similar to
what is commonly associated with magnetic reconnection.
Such scenarios may explain explosive/eruptive phenomena like magnetar giant flares;
like outflows in WD atmospheres.

\section{Acknowledgements}

Authors acknowledge the support from Shota Rustaveli Georgian
National Foundation Grant Project No. FR17-391.
Work of SMM was supported by US DOE Contract No.DE-FG02-04ER54742.


%



%

\appendix

\section{Appendix - Deriving Quadruple Beltrami Equation and Its Analysis}

\bigskip

Plugging Eq. (\ref{B14}) into the Eq.-s (\ref{B13'},\ref{B13}) and then
using them in Eq. (\ref{B9h}), and after some tedious algebra, we get
\begin{equation}
2\eta \,G_0 \nabla \times \nabla \times
\nabla \times {\bf B} \ + \eta \, G_0\ \alpha_1 \
\nabla\times\nabla\times {\bf B} \
+ \ 2\eta \,\alpha_2 \ \nabla \times
{\bf B} - 2 \alpha_3 \ {\bf B} = \label{B-4}
\end{equation}
\[
\hspace{3.5cm} = - \ 2 \alpha \,G_0\,\nabla
\times \nabla \times {\bf V}_h - \alpha\,\beta^{-1}\,G_0^2 \,\alpha_4 \,
\nabla \times {\bf V}_h
+ \ \frac{\alpha}{2}\,\beta^{-1}\,\alpha_5{\bf V}_h \ ,
\]
where
\[
\alpha_1 = a_d[1+\beta^{-1}G_0] + a_i(1+\alpha)\beta \,[1+(1+\alpha)\beta^{-1}G_0] \ , \
\qquad \qquad
\alpha_2
=[1 + (1+\alpha)\,\beta \, ] + \frac{1}{2} (1+\alpha)\ a_d\,a_i \ ,
\]
\begin{equation}
\alpha_3=(a_i-a_d) + \alpha\,(1-\alpha - \beta)\,a_i  \ ,
\qquad \qquad
\alpha_4 = (1 - \beta G_0^{-1}\,[a_d + a_i\,(1+\alpha )\,]\,) \ ,
\ \label{B-5}
\end{equation}
\[
\alpha_5 = [a_d + a_i\,(1 + \alpha )\,\beta \,] + \eta^{-1}\,[a_d - a_i\,(1 + \alpha )\,\beta \,]
- \ 2a_d\,[a_d - a_i\,(1 + \alpha )\,\beta \,] \ .
\]
The equation (\ref{B-4}) with (\ref{B-5}) for no hot electron-fluid fraction
($\alpha \equiv 0$) will eventually give the so called ''Triple Beltrami'' equation
for the magnetic field \ ${\bf B}$ (i.e. l.h.s. of Eq.(\ref{B-4})
$ \equiv 0$) for e-i plasma with degenerate electrons. While for classical
pure e-i plasma we see that $\alpha_1 \to \zeta\,a_d $ and the 2nd term in l.h.s.
of Eq.(\ref{B-4}) can become much bigger than the 1st term ($\zeta \gg 1$)
and one obtains the so called double-curl equation leading to Double Beltrami States.

\bigskip

Using Eq. (\ref{B9h}) in eq. (\ref{B-4}), one obtains for arbitrary $\alpha $
[$H_h\equiv H_{0h}=H_0$, we do not study the heating/cooling problem]
the quadruple Beltrami equation (QB) for ${\bf V}_h$:
\begin{equation}
\eta \,G_0\,\nabla \times \nabla \times \nabla \times \nabla
\times {\bf V}_h + \eta \,G_0 b_1\,\nabla \times \nabla \times
\nabla \times {\bf V}_h
+ \eta \,b_2\,\nabla \times \nabla
\times {\bf V}_h - b_3\,\nabla \times {\bf V}_h -2\alpha \,b_4\,{\bf V}_h =
0 \ , \label{QB-1A}
\end{equation}
where
\[
\qquad \qquad b_1 = \alpha_1 + 2\,\alpha\,a_h \ ;
\qquad \qquad
b_2 = 2\,\alpha\,H_0 +
2\,\alpha \,G_0\,(\alpha_1\,a_h + 2\,\eta^{-1}) \ ;
\]
\begin{equation}
\qquad \qquad
b_3 = 2\,H_0\,\alpha_3 - \alpha \,( \beta^{-1}\,G_0^2\,\alpha_4
+2\,\eta\,\alpha_2\,a_h) \ ; \qquad \qquad b_4 =
\frac{1}{4}\,\beta^{-1}\,\alpha_5 + \alpha_3\,a_h \ . \label{QB-3A}
\end{equation}

It is clear that the large scale is automatically introduced into the system
since \ $b_4\neq 0\ , \alpha \neq 0$ \ in Eq.(\ref{QB-1A}).

Solving the Eq.(\ref{QB-1A}) for ${\bf V}_h$ and plugging it
into (\ref{B9h}) we will get the equation for ${\bf B}$.

\bigskip

Assuming $\alpha \ll 1$ for our problem of study one
can simplify the coefficients in Eq.(\ref{QB-1A}) as follows:
\[
\qquad \qquad \eta = (a_d + \beta \, a_i)^{-1} \ ; \qquad \alpha_1 = (1 + \beta^{-1}\,G_0)\,(a_d + \beta \, a_i) \ ;
\qquad \alpha_2 = \left[(1 + \beta ) + \frac{1}{2}\,a_d\,a_i \right] \ ;
\]
\begin{equation}
\qquad \qquad \alpha_3 = a_i - a_d \ ;
\qquad
\alpha_4 = [1 - \beta \,G_0^{-1}\,(a_i + a_d\,)] \ ;
\qquad
\alpha_5 = [(a_d + \beta \, a_i) - (a_d - \beta \,a_i)^2\,] \
\label{alphas}
\end{equation}
leading to:
\[
\qquad \qquad b_1 = (1 + \beta^{-1}\,G_0)\,a_d + (1 + \beta )\,G_0\, a_i + 2\,\alpha \,a_h \ ;
\]
\[
\qquad \qquad b_2 = 2\,H_0[\,(1 + \beta) + 0.5\,a_d\,a_i\,]
+\alpha \,G_0[\,(1 + \beta^{-1}\,G_0)\,a_d\,a_h +
(G_0 + 3\beta) +2\,a_d\,] \ ;
\]
\[
\qquad \qquad b_3 = 2\,H_0(a_i - a_d) - \alpha \, [\beta^{-1}G_0^2 - G_0\,(a_i + a_d)\,]
- \alpha \,\frac{2a_h}{a_d + \beta a_i}\,[\,(1 + \beta) + 0.5\,a_d\,a_i\,]  \ ;
\]
\[
\qquad \qquad b_4 = \frac{1}{4}\,[a_d\,(1-a_d) + \beta \,a_i\,(1-\beta\,a_i)
+ 2\,\beta \,a_i\,a_d \,]
+ (a_i - a_d)\,a_h \ .
\]

The quadruple Beltrami Equation ({\ref{QB-1A}}) can be factorized as
\begin{equation}
\qquad \qquad (curl - \mu_1)(curl - \mu_2)(curl - \mu_3)(curl - \mu_4)\ {\bf
V}_i = 0 \ , \label{Q-curlA}
\end{equation}
where \ $\mu_i$-s \ define the coefficients in
Eq.(\ref{QB-1}) and are the functions of \ $\alpha , \
\beta , \ n_{0d} , \ H_0$ \ and the degeneracy-determined mass factor \
$G_0$ . The general solution of Eq.(\ref{Q-curl}) is a sum
of four Beltrami fields ${\bf F_k}$ (solutions of Beltrami
Equations $\nabla \times {\bf F}_k = \mu_k {\bf F}$) \ while
eigenvalues ($\mu_k $) of the {\it curl} operator are the
solutions of the fourth order equation
\begin{equation}
\qquad \qquad \mu^4-b_1^{*}\,\mu^3 + b_2^{*}\,\mu^2 - b_3^{*}\,\mu
+ b_4^{*} = 0 \ ,
\label{mu-eqA}
\end{equation}
where
\begin{equation}
\qquad \qquad b_1^{*} = b_1\ ; \qquad b_2^{*} =  G_0^{-1}b_2 \ ; \qquad b_3^{*} = (\eta G_0)^{-1}b_3 \ ;
\qquad
b_4^{*} = 2\,\alpha\,(\eta G_0)^{-1}b_4 \ .
\label{b*A}
\end{equation}

The details of analysis of above equation can be found in \citep{SMB_multi}
for different physical system. Such analysis shows that for a rather big range of
parameters there is a guaranteed scale separation in 2-temperature relativistic e-i
plasma with degenerate (bulk) electrons and small fraction of relativistically
hot electrons.

\end{document}